\documentstyle[12pt]{article}

\def\void{}
\def\labelmark{}

\newenvironment{formula}[1]{\def\labelname{#1}
\ifx\void\labelname\def\junk{\begin{displaymath}}
\else\def\junk{\begin{equation}\label{\labelname}}\fi\junk}%
{\ifx\void\labelname\def\junk{\end{displaymath}}
\else\def\junk{\end{equation}}\fi\junk\labelmark\def\labelname{}}

{\ifx\void\labelname\def\junk{\end{array}\end{displaymath}}
\else\def\junk{\end{array}\right.\end{equation}}
\fi\junk\labelmark\def\labelname{}\def\junk{}
}

\newcommand{\beq}{\begin{formula}}
\newcommand{\eeq}{\end{formula}}
\newcommand{\beqv}{\begin{formula}{}}

\newcommand{\newsection}{
\setcounter{equation}{0}
\section}
\newcommand{\bea}{\begin{eqnarray}}
\newcommand{\eea}{\end{eqnarray}}
\newcommand{\rf}[1]{(\ref{#1})}
\newcommand{\g}{\gamma}
\renewcommand{\l}{\lambda}

\newcommand{\n}{\nu}
\newcommand{\k}{\kappa}
\newcommand{\m}{\mu}
\renewcommand{\d}{\delta}
\newcommand{\th}{\theta}

\newcommand{\sg}{\sigma}

\newcommand{\aR}{\langle R\rangle}
\newcommand{\aRR}{\langle R^2\rangle}
\newcommand{\cD}{{\cal D}}

\newcommand{\cT}{{\cal T}}
\newcommand{\cN}{{\cal N}}
\newcommand{\cL}{{\cal L}}

\newcommand{\noi}{\noindent}

\begin{document}
\topmargin 0pt
\oddsidemargin 5mm
\headheight 0pt
\topskip 0mm

\addtolength{\baselineskip}{0.20\baselineskip}
\hfill NBI-HE-92-53

\hfill July 1992
\begin{center}

\vspace{36pt}
{\large \bf
Quantum gravity, dynamical triangulations
and higher derivative regularization}

\vspace{24pt}

{\sl Jan Ambj\o rn }

\vspace{12pt}

The Niels Bohr Institute\\
Blegdamsvej 17, DK-2100 Copenhagen \O , Denmark\\

\vspace{18pt}

{\sl Jerzy Jurkiewicz}{\footnote {Partly supported by the
KBN grant no. 2 0053 91 01}}    \\

\vspace{12pt}

Institute of Physics, Jagellonian University, \\
ul. Reymonta 4, PL-30 059 Krak\'{o}w 16, Poland.

\vspace{18pt}

{\sl Charlotte F. Kristjansen}

\vspace{12pt}

The Niels Bohr Institute\\
Blegdamsvej 17, DK-2100 Copenhagen \O , Denmark\\

\end{center}

\vfill

\begin{center}
{\bf Abstract}
\end{center}

\vspace{12pt}

\noi
We consider a discrete model of euclidean quantum gravity in four dimensions
based on a summation over random simplicial manifolds. The action used is
the Einstein-Hilbert action plus an $R^2$-term.
The phase diagram as a
function of the bare coupling constants is studied in the search for a
sensible continuum limit. For small values of the coupling constant of the
$R^2$ term the model seems to belong to the same universality class
as the model with pure Einstein-Hilbert action and exhibits the same
phase transition.
The order of the transition may be second or higher. The
average curvature is positive at the phase transition, which makes it
difficult to understand the possible scaling relations of the model.

\vspace{24pt}

\vfill

\newpage

\newsection{Introduction}

Understanding the theory of quantum gravity remains one of the greatest
challenges in theoretical physics. One can try to circumvent the problem
by embedding gravity in a larger theory like string theory. This is in
many respects an appealing approach, but it seems to have lost some of
its momentum and there is not much hope that it will be possible in the
near future to arrive in a natural way to an effective theory of gravity
(and matter) in four dimensions. Discussing basic principles one should
maybe not be so  worried about our technical
inability to deduce the consequences of string theory since this is not
the first time in theoretical physics we are unable to extract
anything but the
simplest perturbative consequences of an otherwise healthy and probably
correct theory. It is more worrisome that string theory is not (yet)
well defined beyond the loop expansion. Again this might not seem
so disastrous since the same was (and to some extent still is) true
for ordinary field theory. However, in the last three years there
{\it has} been a significant progress in our understanding of the
problems connected with the summation of all loops in string
theory. The message has not been encouraging.  At the moment we have
no general principles which allow us to define in an unambiguous way the
summation over all genera in string theory. From this point of view
it might be somewhat premature to announce string theory as the
fundamental theory. Strictly speaking it is not yet a theory but
a set of rules which allow us to calculate certain perturbative
quantities.

If we decide to drop string theory as the theory which will teach us
the nature of quantum gravity it might be (good ?) conservative
policy to stay entirely within the ordinary field theoretical framework. At
first glance it does not look too promising either. The four dimensional
theory is hampered by being non-renormalizable and we do not at present
know any example where such a theory can be defined non-perturbatively,
is non-trivial and at the same time satisfies what we usually view as
the axioms of quantum field theory.  At the same time one of the lessons
of the last thirty years is that field theory is deeply connected to
the theory of critical phenomena via the path integral and has a natural
formulation in Euclidean space-time. But precisely for gravity the
Einstein-Hilbert action is unbounded from below due to the conformal mode
and the Euclidean path integral is ill-defined. One could try to make sense
of the unboundedness of the action either by a contour rotation associated
with the conformal mode as suggested by Hawking and others\cite{hawking}
or by  stochastic regularization
(the so-called fifth time action) as advocated by Greensite and Halpern
\cite{jh,greensite}. It is not the purpose here to
enter into a discussion of the
virtues and drawbacks of these interesting suggestions.  Let us only mention
that they do not really have the flavour of general descriptions
based on ordinary field theory. Of course it is wise to
bear in mind that
if any field theory should depart from basic axiomatic principles
it is quantum gravity, but lacking a general alternative we have
decided to return to analysis of quantum gravity in the context of
ordinary field theory.

Field theory suggests one rather simple minded way out of the above
mentioned problems. This was already discussed long ago by Weinberg who
called it asymptotic safety. The idea is simply that when we, by means
of the renormalization group equations,  work
our way back from the infrared fixed point where the Einstein-Hilbert
action seems a good effective description we will at some point reach
a non-trivial ultraviolet fixed point. In addition the associated
critical surface is assumed to be finite dimensional, which means that
only a finite number of parameters are left arbitrary in the theory, which
from this point of view  can be said not to differ much from ordinary
renormalizable field theories. The effective Lagrangian description of the
theory by means of fields suitable for the infrared fixed point might
then be an infinite series
\beq{*1.1}
\cL_G =\sqrt{g}
\left[ \Lambda -\frac{1}{16\pi G}R + f_2R^2 +f'_2 R_{\m\n}R^{\m\n}+
\cdots \right]
\eeq
which might even be non-polynomial, but which might now (and we will assume
this is the case) make sense if we make a formal rotation to Riemannian
space where the metric has the signature $\{+1,+1,+1,+1\}$
(the generalization of the rotation to Euclidean space in ordinary
field theory).

Of course one weakness in this scenario is
that the existence of the ultraviolet fixed point has been entirely
hypothetical. Further we have not exactly been flooded with examples
of this kind in field theory, as already mentioned. Finally, and we agree
that this point is an annoyance, such a solution does not
have the appeal and aesthetical beauty of the original theory which we want
to quantize. If the quantum theory of gravity offers to us a solution like
\rf{*1.1} the next task must be to find a simpler description
in terms of other variables, maybe somewhat like the  switch in hadronic
physics from hadrons to quarks and gluons.

A few things have happened since Weinberg outlined the above strategy. There
exist now regularizations of the path integral which allow us to define
theories like \rf{*1.1} non-perturbatively and further the progress in
computer science has made it possible to calculate approximately these
path integrals. It is therefore possible to explore, by numerical means,
the phase diagram of the regularized theory and try to locate
phase transition points in the coupling constant space. If the transitions
are of second order one can attempt to define a continuum limit.
The approach has one virtue: it requires only a finite amount of work
to verify whether the idea of a non-trivial ultraviolet fixed point is
viable or not.

The rest of this article is organized as follows: In section 2 we
define the  discretized model to be used to regularize the path integral.
Section 3 provides some details about the numerical method used, while
section 4 describes the results of extensive numerical simulations.
Section 5 contains a discussion of the results obtained so far.

\newsection{The discretized model}

The continuum theory of gravity is reparametrization invariant.
If we discretize the theory in order to regularize it we will
have to break this invariance   provided the action depends on the metric.
An alternative is to consider theories which depend only on topology.
A very interesting approach in this direction in three dimensions
has recently been followed by many people
following the work of Turaev and Viro \cite{topology}
 and it has now been generalized to four dimensions \cite{ooghuri}.
Unfortunately the precise connection with the usual continuum
version of Einstein gravity is not yet clear, especially in four dimensions.

If we restrict ourselves to the conservative approach of
discretizing Einstein's
theory of gravity we will break reparametrization invariance since an action
like \rf{*1.1} depends explicitly on the metric and without having done
anything yet we can already now say that the most important question to
answer in case one manages to carry out successfully the program outlined
in the introduction will be to check that the theory defined by approaching
in a well defined way the ultraviolet fixed point is really reparametrization
invariant. This is by no means obvious since the regularization has broken
this invariance explicitly.

\subsection{Quantization of Regge calculus}
Two rather different regularization schemes have been suggested. The oldest
one goes back to Regge \cite{regge} and we will call it Regge calculus.
It was originally invented as a means to approximate a given smooth manifold
by a piecewise flat Riemannian manifold,
obtained by a suitable triangulation of the
smooth manifold. Regge showed that it was still possible to assign in a
sensible way a concept of curvature to such a piecewise flat manifold. For
$d$-dimensional manifolds the building blocks would  be $d$-dimensional
simplices and the curvature  assigned to the $d-2$-dimensional sub-simplices.
In this way one has both volume and curvature assigned to the piecewise
flat manifold and it is possible to approximate the continuum
Einstein-Hilbert action which reads
\beq{*2.1}
S[g] = \l \int d^d\xi \sqrt{g} - \frac{1}{16\pi G} \int d^d \xi \sqrt{g} R
\eeq
by the following discretized expressions:
\bea
\int d^d\xi \sqrt{g} & \to & \sum_j V_j (d) \label{*2.2} \\
\int d^d\xi \sqrt{g} R & \to & \sum_j V_j(d)
\left [2 \delta_j \frac{V_j ( d-2)}{V_j (d)}\right]. \label{*2.3}
\eea
In \rf{*2.2}-\rf{*2.3} the summation is over
all $d-2$-dimensional sub-simplices  $j$ with volume $V_j(d-2)$.
$\d_j$ is the so-called
deficit angle associated with the $d-2$ dimensional sub-simplex $j$, while
the $d$-dimensional volume $V_j (d)$ associated with the $d-2$
dimensional sub-simplex $j$ is defined as
\beq{*2.4}
V_j(d) = \frac{2}{d(d+1)} \sum_{i \ni j} V_i (d)
\eeq
where the summation is over all $d$-dimensional simplices $i$
which contain the subsimplex $j$.
It is possible to show that by a suitable refinement of the triangulation
of the smooth manifold the discretized expressions \rf{*2.2} and \rf{*2.3}
will actually converge to the continuum value. In this way Regge calculus
provides a geometrical coordinate-independent description of gravity where
it is natural to use the length of the links (the geodesic length
of the edges in the triangulation of the given manifold)
as dynamical variables since they completely specify the flat $d$-simplices
used as building blocks. Originally the method was used mainly in a
classical context where there are no conceptional problems connected with
the approach. However, already as early as in 1968 Regge and Ponzano in an
impressive paper, which contains also the seed to recent development in
topological gravity mentioned above, pointed out that quantum mechanical
amplitudes in three-dimensional Regge calculus can be defined
by a functional integral and maybe computed non-perturbatively. However,
if we seriously want to apply the Regge calculus directly in the
functional integral it looses some of its beauty. In the classical context
the geometry was specified by the length of the links of the building
blocks and the incidence matrix which specified how the building blocks
were glued together. This incidence matrix which determines
the topology was fixed
and not considered a dynamical variable. When we use the Regge formalism
in the path integral the situation is opposite. The task is not to approximate
a given continuum Riemannian manifold but (at least) to sum over
equivalence classes of metrics associated with a given manifold.
Unfortunately there is no simple  one-to-one
correspondence between link length and equivalence classes of metrics, as is
easily seen by considering  triangulations of the two-dimensional plane.
Obviously many triangulations correspond to the same Riemannian  geometry.
This means that a nasty jacobian is involved if we want to use link
lenghts
as our integration variables. In addition one has to choose an integration
measure which ensures that the link lengths satisfy the triangle inequalities
and their higher dimensional analogues, which express that the
$k$-dimensional volume of $k$-dimensional sub-simplices in the given
triangulation must be positive. A great deal of work has gone into
understanding and repairing these shortcomings of conventional Regge
calculus. For a recent excellent review and references we refer to
\cite{hamber}.  While the classical Regge calculus gives a coordinate
independent geometrical description of gravity it of course has nothing
like reparametrization invariance\footnote{One might try to {\it define} the
analogue of local coordinate transformations, see for instance \cite{rw}.}.
It is therefore necessary to prove that this invariance is recovered at
the point in coupling constant space where the continuum limit is taken.
Unfortunately the ``quantum Regge calculus'' has not, in our opinion,
been so successful in this respect. For instance computer simulations in
\cite{gh} seemingly give the wrong coupling to Ising spins in two
dimensions where the coupled spin-gravity system can be solved explicitly
in the continuum. Hopefully this is due to problems with the simulations
rather than a basic flaw in the approach, but we do not know for sure.

\subsection{Dynamical triangulation gravity}

Due to the above mentioned problems with the translation of classical
Regge calculus to quantum theory we  will here use another approach
which has been extensively used in the last few years in the study of
two-dimensional gravity and non-critical strings (which are nothing but
two-dimensional gravity coupled to special matter fields)
\cite{david1,adf,david2,kkm}.  We will call it dynamically triangulated
gravity or (interchangeably) simplicial gravity.

In this approach the fundamental building blocks are regular simplices.
In $d=2$ this means equilateral triangles, in $d=3$ regular tetrahedra.
One now constructs the manifolds by gluing together
the regular $d$-dimensional
simplices  along their $d-1$ dimensional sub-simplices, in such
a way that they form a piecewise flat manifold. The dynamics is  shifted
from  the length of the links to the connectivity of the piecewise linear
manifold and as we shall see there will not be the over-counting present
in the quantum Regge prescription.

The assignment of volume and curvature for a given
triangulation $T$ created by gluing together the regular simplices
is in this case very simple. Let us introduce the following notation:
An $i$-dimensional (sub)-simplex  is denoted $n_i$, i.e. vertices
are denoted $n_0$, links $n_1$ etc. The total number
of such (sub)-simplices in $T$ is denoted $N_i (T)$. By the order
$o(n_i)$ we understand the number of $d$-dimensional simplices which
share the sub-simplex $n_i$. We will usually consider only  the class of
regular simplicial manifolds where we have put the following
restrictions on $o(n_i)$:
\beq{*2.5}
  o(n_{d-1})=2,~~~~~o(n_i) \geq d-i+1,~~~~~(i \leq d-2).
\label{1.a0}
\eeq
\rf{*2.2}-\rf{*2.3} reduce to
\beq{*2.6}
\int d^d\xi \sqrt{g} \propto   N_d
\eeq
and
\beq{*2.7}
\int d^d\xi \sqrt{g} R \propto  \sum_{n_{d-2}} (c_d- o(n_{d-2})).
\eeq
The constant $c_d$ in \rf{*2.7} should be adjusted in such a way that
for
a hypothetical triangulation of flat space the sum should give zero. For $d=2$
one can triangulate flat space with regular triangles. The order of
each vertex is 6 and consequently $c_2=6$. Higher dimensional flat
space does not admit a regular tessellation, but one can still ask
for the average value of $o(n_{d-2})$ required to fill up $d$-dimensional
flat space. The angle $\th_d$ between two $d-1$ dimensional simplices
belonging to the same $d$-simplex is given by
\beq{*2.8}
\cos \th_d = 1/d
\eeq
and in order to fill up $d$-dimensional space we need to have
\beq{*2.9}
o(n_{d-2})= 2\pi/\th_d \equiv c_d
\eeq
This is the constant which enters in \rf{*2.7}.
We find
\beq{*2.10}
c_2= 6,~~~~~c_3=5.104,~~~~~c_4=4.767.
\eeq
Let us further note that
\beq{*2.11}
\sum_{n_{d-2}} o(n_{d-2}) = \frac{(d+1)d}{2} N_d
\eeq
since there are
$\left(\begin{array}{c} d+1 \\ d-1 \end{array} \right)$  $d-2$ dimensional
sub-simplices in a $d$-dimensional simplex. The discretized version
of the continuum action can now, for a given triangulation $T$, be
written as
\beq{*2.12}
S_d [T] = \k_d N_d(T) -\k_{d-2} N_{d-2}(T).
\eeq

A more general action would be the following:
\beq{*2.13}
S_d [T] = \sum_{i=0}^d \k_i N_i(T)
\eeq
involving the fugacities for all different $i$-dimensional
sub-simplices. Of course one can choose to consider actions which
cannot be expressed entirely in terms of the $N_i$'s. The higher derivative
terms which can be added to the continuum action \rf{*2.1}
will in general be of this kind, and we are going to consider them
later, but let us for the moment discard such terms.
Not all $N_i$'s are independent.
The relations between the $N_i$'s can be worked out by the requirement
that the triangulation should be locally homeomorphic to $R^d$. This means
for instance that all $n_d$'s sharing a given vertex $n_0$ should be
homeomorphic to the unit ball in $R^d$. Similar restrictions hold for
the neighbours to an $i$-dimensional simplex $n_i$ in the triangulation,
and the relations this imposes on the $N_i$'s are summarized in
the so-called Dehn-Sommerville\cite{ds} relations
\beq{*2.14}
N_i = \sum_{k=i}^d (-1)^{k+d}
\left(\begin{array}{c} k+1 \\ i+1 \end{array} \right)  N_k,
\eeq
valid for all $i \geq 0$. These relations are not independent, but allow us
to eliminate all $N_{2i+1}$'s if $d$ is even and all $N_{2i}$'s
if $d$ is odd. In the case of even dimensions we have for a given
triangulation  $T$ in addition Euler's relation
\beq{*2.15}
\sum_{i=0}^d (-1)^{i+d} N_i(T) = \chi_d(T).
\eeq
where $\chi_d(T)$ denotes the Euler characteristic of the piecewise
linear manifold which corresponds to the triangulation $T$.
Of course this relation is only useful if we know the Euler characteristic
of the triangulation $T$. As we shall see the restriction of topology
is very important, and in case we fix the topology of $T$ we can
use \rf{*2.15} to eliminate for instance $N_0$. In case the topology is
not fixed we can trade $N_0(T)$ for $\chi_d(T)$.  For odd dimensions
\rf{*2.15} follows from the Dehn-Sommerville relations with
\beq{*2.16}
\chi_{d=2n+1} = 0.
\eeq
In odd dimensions the Euler characteristic is identically zero for
any simplicial manifold, and it follows just from the requirement
of local homeomorphism to $R^d$.

The recipe for going from the continuum functional integral to
the discretized one is now:
\bea
\int \cD [g_{\m\n}]& \to &
\sum_{T \sim {\cT}} \label{*2.17} \\
\int \cD [g_{\m\n}]\;e^{-S[g]} & \to &
\sum_{T \sim {\cT}} e^{-S[T]} \label{*2.18}
\eea
The formal integrations on the lhs of \rf{*2.17} and
 \rf{*2.18} are over all equivalence classes of metrics, i.e the volume
of the diffeomorphism group is divided out. $\cT$ denotes a suitable
class of triangulations. One class of constraints is given by \rf{*2.5},
but it should be stressed that such short distance restrictions are
not expected to be important in the scaling limit.

Since different triangulations give rise to different curvature
assignment one can view the above summation as a summation over
different Riemannian manifolds. There is no problem with over-counting
in this formulation. The idea of the continuum functional
integral is precisely to perform such a sum with weight $e^{-S[g]}$.
Of course the discretized sum on the rhs of \rf{*2.17} and \rf{*2.18}
can only be viewed as an approximation to the continuum expression
which hopefully ``converges'' in the scaling limit to the correct
expression. The questions which are difficult to answer are whether
the class of piecewise flat manifolds is ``close'' to the class
 of Riemannian manifolds and whether the piecewise linar manifolds
 are selected
sufficiently uniformly with respect to Riemannian manifolds that
\rf{*2.17} and \rf{*2.18} are good approximations. Unfortunately there
is no weak coupling expansion where one can check this, but it is
very encouraging that the formalism is known to work in the
two-dimensional case, even if one couples conformal matter with central
charge $c \leq 1$ to the system. In this case it is possible to
solve both the continuum and the discretized system. In particular we see
that reparametrization invariance is recovered in the scaling limit.

Since $d=4$ has our main interest in this work we can
 write our partition function as
\beq{*2.19}
Z(\k_2,\k_4) = \sum_{T \in \cT} e^{-\k_4 N_4 + \k_2 N_2}
\eeq
This is the grand canonical partition function, where the volume of
the universe can vary. It is sometimes convenient to change from the
grand canonical ensemble described by \rf{*2.19} to the canonical
ensemble where the volume $N_4$ is kept fixed. The corresponding
partition function will be
\bea
Z(\k_2,N_4)& = &\sum_{T \sim \cT (N_4)} \;  e^{\k_2 N_2(T)}
\label{*2.20} \\
Z(\k_2,\k_4)& =& \sum_{N_4} \; Z(\k_2,N_4)\;e^{-\k_4 N_4}. \label{*2.21}
\eea

If the entropy, i.e. the number of configurations for a given $N_4$,
is exponentially bounded it is easy to prove that there is a
critical line $\k_4=\k_4^c (\k_2)$ in the $(\k_2,\k_4)$-coupling
constant plane. For a given $\k_2$ the partition function
\rf{*2.19} will then be well defined for $\k_4 > \k_4^c(\k_2)$. Let us call
this domain in the coupling constant plane $\cD$. Critical behaviour
can be found only when we approach the boundary $\partial \cD$ which we
denote the critical line, but in general we only expect interesting
critical behaviour at certain critical points on $\partial \cD$
(i.e. at certain values of $\k_2$). These are the points we are looking
for in the numerical simulations.

Let us end this subsection by discussing a point which is worth emphasizing.
We have been deliberately vague defining the class of triangulations
$\cT (N_4)$ over which the summation is to be performed in  a formula
like \rf{*2.20}. Already in two dimensions where the classification
of topology is so simple (it is defined by the Euler number $\chi$)
an unrestricted summation over manifolds of different topology does cause
problems. In fact the two-dimensional analogue of \rf{*2.20} and \rf{*2.21}
which by means of Euler's relation can be written:
\bea
Z(\tilde{\k}_0,N_2)& = &\sum_{T \sim \cT (N_2)} e^{\tilde{\k}_0 \chi}
\label{*2.22} \\
Z(\tilde{\k}_0,\tilde{\k}_2)&=& \sum_{N_2} Z(\tilde{\k}_0,N_2)
e^{-\tilde{\k}_2 N_2}  \label{*2.23}
\eea
does not make any sense. The well known reason is that the number of
triangulations, $\cN (N_2)$, which
one can make by gluing together a given number
$N_2$ of equilateral triangles to a two-dimensional surface is too
large. It grows factorially fast: $\cN (N_2) \geq N_2 !$. This means that
the two-dimensional analogue of \rf{*2.21} is never convergent. This is
not a spurious result of a perverse discretization. An analogous result
has been proven in the continuum two-dimensional theory where the volume
of moduli space grows at least factorially with the genus \cite{gp}.
It is the same effect we observe in the discretized case: As long as
we fix the topology, i.e. the Euler characteristic $\chi$ of the surface,
we have: $\cN_\chi (N_2) \sim N_2^{\g_\chi -3} \exp (\tilde{\k}_{2c} N_2)$,
i.e. only
an exponential growth of the number of surfaces. In this case \rf{*2.23}
is well defined for a certain range of $\tilde{\k_2}$'s. However,
an unrestricted
summation over topologies makes the subleading pre-exponential factor
$N_2^{\g_\chi}$ dominant when $|\chi| \sim N_2$ since $\g_\chi \sim -5\chi/4$.

In higher dimensions the situation is of course only worse and the best we
can hope for is a well defined expression for a fixed (or at least restricted)
topology. In the
following we will restrict ourselves to four-dimensional  manifolds
with the topology of $S^4$.

The above outlined non-perturbative definition of gravity has nothing
to add to our understanding (or rather lack or understanding) of the
question of whether or not to sum over topologies in quantum gravity.
Apart from the problem that the topologies of  non-simply connected
four-dimensional manifolds  cannot be classified in a sensible way, the
partitionfunction does not even make sense if we only restrict ourselves to
the sub-class of topologies which one can construct by simple analogy
to two-dimensional surfaces of genus $g$. In the rest of this article
we will only be interested in the search  for  non-trivial fixed
point of the above defined theory (modified with higher curvature terms)
where the class of triangulations $\cT$ corresponds to manifolds
with the topology
of $S^4$.

\subsection{Higher curvature terms}

There is no straightforward generalization of Regge's work to
theories of gravity which involve higher derivative terms like
$R^2$ in the action. The reason is that Regge viewed the piecewise
flat manifold, not as
a discrete approximation to  an underlying continuum surface,
but as one where curvature could be defined
in a mathematical stringent way, entirely
in terms of the geometrical concepts involved in parallel transportation.
In this way the curvature occurs in $\d$-functions on the lattice
geometry, with support on the $d-2$-dimensional sub-simplices.  Anything
other than the Einstein action (and the cosmological term) will then
involve higher powers of $\d$-functions and this means that for piecewise
flat manifolds, interpreted  as by Regge, terms like $\int \sqrt{g} R^2$
{\it are} infinite. In order to make sense of higher derivative terms
one has to change the perspective on Regge calculus somewhat as advocated
by Hamber and Williams \cite{hw1} and view the
lattice geometry as representing
an approximation to some smooth geometry and the local curvature
as some average curvature for a small volume. In fact our formulas for
Regge calculus have already hinted this interpretation in the sense
that we have assigned a volume density $V_{n_{d-2}}(d)$
 to each $d-2$-dimensional sub-simplex
$n_{d-2}$ which can be viewed as an appropriate share of the volumes of the
$d$-dimensional simplices to which the sub-simplex $n_{d-2}$ belongs.
In the same way we have written the curvature density $R$ as
$\d_{n_{d-2}}/V_{n_{d-2}}(d)$ viewing it formally as representing some
average value in the volume $V_{n_{d-2}}(d)$. With such an interpretation
one can of course write
\beq{*2.23a}
\int d^d\xi \sqrt{g} R^2  \sim \sum_{n_{d-2}} V_{n_{d-2}}(d)
\left[ \frac{ 2\d_{n_{d-2}}\,V_{n_{d-2}}(d-2)}{V_{n_{d-2}}(d)} \right]^2
\eeq
This definition must be interpreted with some care if we want convergence
to the  continuum  value for a smooth manifold by  successive subdivision
\cite{eliezer}. We do not have to worry too much about these subtleties
here since our task in the functional integral is not to approximate
a given smooth manifold but to select some class of manifolds which
can be used in an (approximate) evaluation of the integral. From
this point of view we will use the $R^2$ term as representing
typical higher derivative terms which one would have to insert in order
to stabilize the Euclidean path integral as explained in the
introduction. As is well known from for instance lattice gauge theories
discretized versions of higher derivative terms are by no means universal.
It is clear that this point of view is not as beautiful as the
original geometrical way that Regge viewed piecewise
flat manifolds, but our perspective is  that a term like
\rf{*2.23a} will probe a universality class of  theories which have an
effective expansion in terms of higher derivative actions like \rf{*2.23a}.

In the case where we consider  the piecewise flat manifolds which can be
obtained by the process of dynamical triangulation as described above
\rf{*2.23a} simplifies and we get
\beq{*2.24}
\int d^d\xi \sqrt{g} R^2 \sim \sum_{n_{d-2}} o(n_{d-2})
\left[ \frac{c_d -o(n_{d-2})}{o(n_{d-2})}\right]^2
\eeq
This formula has a slight problem with the continuum interpretation
since flat space does not have a regular tessellation except for
$d=2$. This means that the term can never scale to zero:
\beq{*2.24a}
\sum_{n_{d-2}} o(n_{d-2})
\left[ \frac{c_d -o(n_{d-2})}{o(n_{d-2})}\right]^2  \geq const. \;\; N_d.
\eeq
If we introduce a scaling parameter $a$, which is to be identified with
the link length, and which is going to be scaled to zero, the physical
volume  $V\equiv\int d^4\xi \sqrt{g} $ being kept fixed, and if we
assume canonical
scaling of the terms involved of both sides of \rf{*2.24a} we get
(restricting ourselves to four dimensions which have our main interest):
\beq{*2.24b}
\left. \int d^4\xi \sqrt{g} R^2 \;\right|_{DT} >
\frac{const.}{a^4} \left.\int d^4\xi \sqrt{g} \;\right|_{DT}.
\eeq
This means that the leading term on the lhs \rf{*2.24a} is just a
cosmological constant term and by expanding the lhs we see that
it also contains an Einstein-Hilbert term etc.. Under the assumption of
naive scaling we have to write instead of \rf{*2.24}
\beq{*2.24c}
 \sum_{n_{d-2}} o(n_{d-2}) \left[ \frac{c_d -o(n_{d-2})}{o(n_{d-2})}\right]^2
\sim
\int d^d\xi \sqrt{g} \left[\frac{c_0}{a^4}+ \frac{c_1}{a^2}R + c_2 R^2
\cdots \;\right].
\eeq
Our lattice ``$R^2$''-term is thus to be considered as a generalized
higher derivative term which, when added to the lattice Einstein-Hilbert
term, in addition will lead to a redefinition of the bare cosmological
coupling constant and the bare gravitational coupling constant.

One interesting aspect of the dynamical triangulation approach is that
for a finite lattice volume it automatically provides a  cut off for the
Einstein action. This is not the case for the conventional Regge calculus
where the action can go to infinity without the volume diverging. The reason
is that the volume (for $d >2$)
of the $d-2$-dimensional sub-simplices can diverge without
the corresponding volume of the $d$-dimensional simplices going to infinity.
In the case of dynamical triangulations we have (restricting again
ourselves to four dimensions)
\beq{*2.25}
-const. \cdot N_4 <\sum_{n_2} ( c_4- o(n_2)) < const. \cdot N_4.
\eeq
{\it If} we assume a conventional scaling in the tentative continuum limit
we can rewrite \rf{*2.25} as
\beq{*2.26}
\left| \int d^4\xi \sqrt{g} R \right|_{DT} \leq
\frac{const.}{a^2} \;\;\left.\int d^4\xi \sqrt{g} \right|_{DT}
\eeq
where  again we have have introduced the link length  $a$,
which is going to be scaled to zero while  the physical
volume  $V\equiv\int d^4\xi \sqrt{g} $ is kept fixed.

\subsection{Observables}

Let us here discuss the observables (see also \cite{adj}
for a more general discussion). Due to diffeomorphism invariance
and the fact that we in quantum gravity have to integrate over all
Riemannian manifolds the  observables which are most readily
available are averages of invariant local
operators like the curvature $R(x)$ and suitable contractions
of powers of the curvature tensor, like $ R_{\m\n}^2$ or $ R_{\m\n\l\sg}^2$,
and the fluctuation
of these averages. In addition one can discuss and measure so-called fractal
properties of space-time and finally with some effort define the
concept of correlators of local invariant operators.

The simplest observable is the average curvature.
If we consider the discretized partition function we have
\beq{*2.40}
\int d^4 \xi \sqrt{g(\xi)} R(\xi) \propto
\frac{2}{d(d+1)} \sum_{n_2}o(n_2) \frac{ c_4-o(n_2)}{o(n_2)}
= \frac{c_4}{10}N_2- N_4
\eeq
and since the volume is
\beq{*2.40a}
\int d^4 \xi\sqrt{g(\xi)} \propto  \frac{2}{d(d+1)} \sum_{n_2} o(n_2) = N_4
\eeq
we can define the average curvature per volume as
\beq{*2.40b}
\aR = \frac{\int d^4 \xi \sqrt{g(\xi)} R(\xi)}{\int d^4 \xi\sqrt{g(\xi)}}
 \propto
\frac{c_4}{10}\frac{ N_2}{N_4}- 1
\eeq
In \rf{*2.40b} $\aR$ is defined for a single manifold. We get of
course the quantum version by calculating the functional average
of $\aR$ over all Riemannian manifolds, weighted by $e^{-S}$. This
is what we will do numerically.
The average curvature is a bulk quantity which will allow us to get
a quick survey of the phase diagram of four-dimensional quantum
gravity.  In a similar way we can define (with the drawbacks described
in the last subsection) $\aRR$ by
\beq{*2.40c}
\aRR = \frac{\int d^4 \xi \sqrt{g(\xi)} R^2(\xi)}{\int d^4
\xi\sqrt{g(\xi)}}
\propto
\frac{\sum_{n_2} o(n_2) \left[(c_4 -o(n_2))/o(n_2)\right]^2}{10N_4}.
\eeq
Again this average is defined over a single manifold and we have eventually to
take the weighted average over all manifolds in our ensemble.
A quantity which will have our interest will be $\aRR -\aR^2$.

A more refined, but related observable is the
integrated curvature-curvature correlation.
In a continuum formulation it would be
\beq{*2.41}
 \chi (\k_2) \equiv \langle \int d^d \xi_1 \; d^d \xi_2 \; \sqrt{g(\xi_1)}
R(\xi_1)
\;\sqrt{g(\xi_2)} R(\xi_2) \;\rangle-\langle\int d^d \xi
\sqrt{g(\xi)}R(\xi)\rangle^2.
\eeq
In a lattice regularized theory one would expect that away from the critical
points short range fluctuations will prevail, while approaching
the critical point long range fluctuation might be important and would
result in an increase in $\chi (\k_2)$. The observable
$\chi (\k_2)$ is the second
derivative of the free energy $F= -\ln Z$ with respect to the gravitational
coupling constant $G^{-1}$.
In the case where the volume $N_4$ is kept fixed we see that
\beq{*2.42}
\chi (\k_2,N_4) \sim \langle N_2^2\rangle_{N_4}-\langle N_2\rangle_{N_4}^2
{}~=~ - \frac{d^2 \ln Z(\k_2,N_4)}{d \k_2^2}
\eeq
{}From the above discussion we have to look  for points
along the critical line $\partial \cD$ where $\chi (\k_2,N_4)/N_4$
diverges in the infinite volume limit $N_4 \to \infty$.

Another observable is the Hausdorff dimension. One can define the
Hausdorff dimension in a number of ways, which are not necessarily
equivalent. Here we will simply measure the average volume
$V(r)$ contained within a radius $r$ from a given point.
In \cite{adj} the concept of a {\it cosmological Hausdorff
dimension} $d_{ch}$ was defined. It essentially denotes the power
which relates the average radius of the ensemble of universes of
a fixed volume to this volume:
\beq{*2.43}
\langle {\rm Radius}\rangle_{N_4}~ \sim~ N_4^{1/d_{ch}}.
\eeq
{}From the distribution $V(r)$ we can try to extract $d_{ch}$. If
 for large $r$ we have the behaviour
\beq{*2.44}
V(r)~ \sim~ r^{d_h}
\eeq
we can identify $d_{ch}$ and $d_h$.

By the use of Regge calculus it is straightforward to convert these
continuum formulas to our piecewise flat manifolds. Between two points
in the piecewise flat manifold there is a geodesic which is a piecewise
linear path. Rather than using this definition we will use approximations
which are much more convenient from a numerical point of view and
which one expects should be sufficient for the purpose of extracting
general scaling behaviour and fractal properties: We define the distance
between two vertices as the shortest path along links which connects
the two vertices, i.e. it is essentially the number of links of
the shortest path since all links have the same length. We call this
distance the ``$n_1$''-distance between vertices and denote its value by $d_1$.
In the same way we can define a shortest path between two 4-simplices as the
shortest path obtained by moving between the centers of neighbour
4-simplices. We call this distance the ``$n_4$''-distance between 4-simplices
and denote its value by $d_4$. The dual graph to a given triangulation,
obtained by connecting the centers of neighbour 4-simplices,
will be a $\phi^5$-graph and the $n_4$-distance on the triangulation will be
the $n_1$-distance on the dual $\phi^5$-graph. A priori these distances
are not related and it is easy to find  triangulations where they
differ vastly  for specific choices of vertices and associated 4-simplices.
But for averages over vertices and over triangulations one would expect
that they carry the same information about the geometry and we shall see
that this is indeed the case.

With the $n_1$- and $n_4$-definitions of geodesic distances on the
triangulations  it becomes trivial to measure numerically  relations
like \rf{*2.44}. As we shall see it is less trivial to extract in a
reliable way the Hausdorff dimension $d_h$.

Let us finally discuss the measurements of correlation functions in
quantum gravity. If we have local invariant operators $O_1(x)$ and $O_2(x)$
(like for instance $R(x)$)
we can define a correlation function of geodesic distance $d$ by
\beq{*2.45}
\tilde{G}(d) =  \int\int  d^4 \xi_1 \sqrt{g(\xi_1)} d^4 \xi_2 \sqrt{g(\xi_2)}
     O_1(\xi_1)O_2(\xi_2) \d ( d(\xi_1,\xi_2) -d).
\eeq
In this definition  $d(\xi_1,\xi_2)$ denotes the geodesic distance
between $\xi_1$ and $\xi_2$ for a given manifold, but \rf{*2.45} makes
sense also as a functional average, and therefore in principle in
quantum gravity. When we discuss numerical data we will always
have in mind the functional average.
It might be convenient to divide by a
volume element to get the dimension of  a point-point correlation function.
If $V(r)$ denotes the volume inside a ``ball'' of radius $r$ we can write
\beqv
dV(r)= V'(r)dr
\eeq
and we define
\beq{*2.46}
G(d) = \tilde{G}(d)/V'(d).
\eeq
In the case where we have a finite Hausdorff dimension the exponential
fall off of $G(d)$ and $\tilde{G}(d)$ are identical, but it needs not be
the case if we have an infinite Hausdorff dimension.

\newsection{The numerical method}

Unfortunately the analytic methods of two dimensional simplicial
gravity have not yet been extended to higher dimensions. The numerical
method of ``grand canonical'' Monte Carlo simulation, which is well
tested in two dimensions \cite{jkp,adfo,afkp},
has recently been applied to three dimensions \cite{av1,am1,bk,av2,abkv}
and in four dimensions \cite{am2,aj,varsted,am3}
A necessary ingredient for Monte Carlo simulations in simplicial
quantum gravity  is a set of so-called ``moves'', i.e. local changes
of the triangulations, which are ergodic in the class of triangulations
we consider.  A general set of moves in any dimension has been known for
a long time \cite{alexandre}. They are however not optimal for numerical
simulations. A more  convenient set of moves
for higher dimensional gravity was suggested
in \cite{am1}. The ergodicity of these moves in three dimensions
was proved in \cite{bk}, and the generalization of
this proof to $d=4$ was given in
\cite{gv}. In $d$ dimensions there are $d+1$ moves.
Their general description is as follows: remove an $i$-dimensional
simplex of order $d+1-i$ and the higher dimensional simplices of which it is
part, and replace it by a $d-i$-dimensional simplex (``orthogonal'' to
the removed $i$-dimensional simplex) plus the appropriate
higher dimensional simplices such that we still have a triangulation.
Such moves will be allowed provided they do not violate
\rf{*2.5} and provided they do not create simplices already present,
i.e which already have the same vertices as the newly created simplices.

Let us
consider four dimensions where there  are five moves.
The first move consists of removing a four-dimensional simplex $n_4(old)$
and inserting  a vertex $n_0(new)$
in the  void interior and adding links (and the induced
higher dimensional simplices) which connect $n_0(new)$
to the five vertices  of $n_4(old)$. In this way $n_4(old)$ is replaced by
five new $n_4$'s and the total change of $N_i$'s is
\beq{*x2.1}
\Delta N_0=1,~~\Delta N_1=5,~~\Delta N_2=10,~~\Delta N_3=10,~~\Delta N_4=4.
\eeq
The second  move consists of removing a three-dimensional simplex $n_3$ and
the two $n_4$'s sharing it, and then inserting a link (``orthogonal'' to
$n_3$) and associated $n_2$'s, $n_3$'s and $n_4$'s.  The total change
of $N_i$'s is in this case
\beq{*x2.2}
\Delta N_0 =0,~~\Delta N_1=1,~~\Delta N_2=4,~~\Delta N_3=5,~~\Delta N_4=2.
\eeq
The third move is a ``self-dual'' move where $\Delta N_i =0$. It consist of
removing a triangle of order three
and associated higher dimensional simplices and
inserting the ``orthogonal'' triangle and its associated higher dimensional
simplices such that we still have a triangulation. The fourth move is
the inverse of the second move, while the fifth move is the inverse
of the first.

The change in the action induced by these moves  can now readily be
calculated and we are in a position to use the standard Metropolis updating
procedure. The weights required for detailed balance are easily determined.
Let us only remark here that the nature of the problem naturally suggests
to use indirect addressing by pointers since  there is no rigid lattice
structure. In addition we found it most efficient to keep pointers
to vertices of order five, links of order four and triangles of order
three since these are the ones used in the updating. Since programs
of the above nature are not well suited for vectorization  it is
optimal to run them on fast workstations.

Since we are forced to use a grand canonical updating where the
volume of the universe $N_4$ is changing, it is convenient to use the
technique first introduced in \cite{bb} and used successfully in
the simulations in three-dimensional gravity \cite{av1,av2}. It allows
us to get as close to a canonical simulation as possible and it
provides at the same time an estimate of the critical point $\k_4^c(\k_2)$ for
a given value of the coupling constant $\k_2$. The idea is the following:
Assume we want to perform a measurement at some fixed value $N_4(F)$.
The task is to constrain the fluctuations of $N_4$ to the neighbourhood
of $N_4(F)$ without violating ergodicity. First we make an
approximate estimate of the critical point $\k_4^c$, which
we denote $\k_4^c(N_4(F))$. It can in principle depend on $N_4(F)$. Next
we choose the the actual $\k_4$ used in the simulation as a function
of the value of $N_4$ in the following way:
\beq{*x2.4}
\k_4(N_4) = \left\{
\begin{array}{lcl}
\k_4^c(N_4(F))- \Delta \k_4 & {\rm ~~~for~~~} & N_4 < N_4(F) \\
\k_4^c(N_4(F))+ \Delta \k_4 & {\rm ~~~for~~~} & N_4 > N_4(F).
\end{array} \right.
\eeq
For sufficiently large values of $N_4(F)$ and small values of $\Delta \k_4$
we will get an exponential  distribution of $N_4$'s peaked at $N_4(F)$:
\beq{*x2.5}
P(N_4) \sim e^{(N_4-N_4(F)) (\k_4^c-\k_4(N_4))}.
\eeq
By monitoring $\Delta \k_4$ we can effectively control the width of
the distribution of $N_4$ around $N_4(F)$ without violating the principle
of ergodicity. A measurement of the exponential distribution also
allows us to determine $\k_4^c$. If the exponentially fall off is different
above and below $N_4(F)$ it means that $\k_4^c$ is different from
our guess $\k_4^c(N_4(F))$ and we can use the optimal value
in the next run. Further the measurements of $\k_4^c$ for different values
of $N_4(F)$ allow us to extract the subleading correction to the
distribution. Assume that the partition function has the form:
\beq{*x2.6}
Z(\k_2,\k_4)= \sum_{N_4} Z(\k_2,N_4) e^{-\k_4 N_4}
\eeq
where
\beq{*x2.7}
Z(\k_2,N_4) \sim N_4^{\g(\k_2)-2} e^{\k_4^c(\k_2) N_4}\;
\left( 1 + O(1/N_4) \right).
\eeq
By our method the critical point $\k_4^c$ determined by measurements
in the neighbourhood of various $N_4(F)$'s would lead to:
\beq{*x2.8}
\k_4^c(measurement) = \k_4^c(\k_2) +\frac{\g(\k_2)-2}{N_4(F)}
\eeq
and a determination of the entropy exponent $\g(\k_2)$, which in fact
governs the volume fluctuations of the system.

\newsection{Numerical results}

The measurements of average curvature etc. were performed for different
values of $N_4$: 4000, 9000, 16000 and 32000. The number of attempted
updatings were of the order $10,000\times N_4$ (sometimes considerably
longer at critical
points where thermalization was slow). The phase diagram was scanned
varying the (inverse) bare gravitational coupling constant
$\k_2$ and the ``$R^2$'' coupling constant which
we denote $h$. For each value of $\k_2$ and $h$ and each value of $N_4$
this implies a fine-tuning of the value of the bare ``cosmological'' coupling
constant $\k_4$ to its critical value $\k_4^c(\k_2,h,N_4)$.
As explained in the last section
it is convenient to perform
the simulations in the neighbourhood of some fixed value of the
volume so we choose a specific $N_4(F)$ and limit the fluctuations in
volume  to  some neighbourhood of $N_4(F)$. If we decide to perform the
measurements after a given number, $n$, of Monte Carlo sweeps, in
practise we
perform the actual measurement the first time, after the $n$'th sweep,
the system passes a state where the value of $N_4$ is equal to $N_4(F)$.
This is what we mean when we say that the measurements
were performed for a given value of $N_4$.

The result for the
average curvature is shown in fig.\ 1 for $h=0$ over a large
range of $\k_2$, while fig.\ 2 shows the average curvature for different values
of $h$, ranging from $h=0$ to $h=20$ and $N_4 = 16000$.
It is not possible with our
choice of $R^2$ term to increase $h$ further since the acceptance rates
in the Metropolis updating  become too small.

We observe the following: In case there is no coupling constant
except the cosmological coupling constant the average curvature
$\langle R  \rangle$ is zero, which is a nice result since it shows
that the selection of manifolds in the context of dynamical triangulations
has no bias towards positive or negative curvature. In case we take
$\k_2$ positive the average curvature will be positive (this corresponds
to the conventional sign of the gravitational coupling constant).
If we take $\k_2$ negative (``anti-gravity'') the average curvature
will be negative.

For $h=0$ and $\k_2 \approx 1.1$ we see a change towards
large positive curvature. The same change is seen for  $h > 0$,
only we have to go to  larger values of $\k_2$ when $h > 10$. For a fixed
positive value of $\k_2$ the curvature decreases with increasing
$h$ as expected.
However, as discussed above, the limit $h \to \infty$ does not really
correspond to zero local curvature due to the special form of the
our discretized ``$R^2$''-term. The absolute minimum value of the
discretized term corresponds to a constant negative curvature:
$R= -0.046$.

In fig.\ 3 we have shown the susceptibility defined by \rf{*2.42}.
It can be measured directly, or as
the derivative of the average curvature.
We have used the second method,
but have also measured the susceptibility directly, with comparable results.
One sees a clear peak which grows somewhat with volume. This
could be taken as a sign that the system becomes critical in this region,
although the system size is not big enough to exclude the possibility
of a phase transition of higher order.

An independent signal of criticality
is found by
looking at the observable $\aRR -\aR^2$. In fig.\ 4a we show
its behaviour as a function of $\k_2$ for $h$ between $0$ and $20$. For small
$h$ values we see a clear peak with a volume dependence
in the region where the $\k_2$ susceptibility has a peak too.
We note the clear asymmetry between the two sides of the peak, especially
for $h=0$ (fig.\ 4b). This explains why the position of the peak seems to be
shifted towards smaller values of $\k_2$ when compared to the susceptibility
curve. Here again our system is not big enough to exclude the possibility
that the increase with  volume is only a finite size effect and that
eventually we shall observe only a discontinuity in the derivative at
a critical point, which again could signal a phase transition of higher
(perhaps 3rd) order.
The signal deteriorates somewhat for large values of $h$,
contrary to the susceptibility signal.
We can draw a critical line in the $(\k_2,h)$-coupling constant plane,
fig.\ 5, and fig.\ 6 shows the average curvature at the transition
point as a function of $h$ for $N_4 = 4000$, $9000$ and $16000$.
We see that the average curvature
gets smaller when $h$ increases and also when $N_4$ increases,
{\it but for all the values we have been able to
probe we have}
\beq{*3.1}
\aR_c \equiv \langle {R(\k_2^c,h)\rangle} > 0
\eeq
{\it and we can not conclude that $\aR_c >0$ is a finite volume effect.}
The data indicate rather that $\aR_c$ remains positive even for $h \to \infty$.
For $h > 12$ there seems to be a qualitative
change in the distributions, which might indicate a different transition,
but we have not found that it cured the problems of the $h=0$ situation.

The computer simulations in three dimensions revealed a similar
situation: A transition and a $\aR_c > 0$.
In three dimensions there was a very
strong hysteresis in the same transition, favouring  a first
order transition. Here we have not seen the same strong hysteresis.
For $N_4=4000$ there was no problem moving from one phase to the other.
For larger $N_4$ we have observed very slow thermalization
and huge fluctuations in geometry close to $\k_2=\k_{2c}$, but it did not
present itself as clear hysteresis.

Let us now explore the change in geometry along the critical line.
Above we have defined the  ``geodesic'' link distance $d_1$ between
vertices and the four-simplex distance $d_4$ between four-simplices.
The average values describe typical radii of our universes. They are shown
in fig.\ 7 ($\langle d_1\rangle$) and fig.\ 8 ($\langle d_4\rangle$).
Although the $d_4$ distances
are approximately six times larger that the $d_1$ distances they clearly
behave qualitatively in the same way and reveal a drastic change in the
geometry as we pass the critical region of  $\k_2$. The nature of the
change seems to be independent of $h$.

The typical universes generated by the computer simulations
have  small radii, almost independent of the volume if we are
below the critical $\k_2$ region. After we have passed the critical
region the radii become quite large and show a very clear volume
dependence. In fact it seems as if the radius grows almost linearly
with volume. {\it Qualitatively this implies that the Hausdorff
dimension is large below the critical region and small (in fact close
to one) above the critical region}. We have not attempted to determine
the larger Hausdorff dimension. The growth in radius with volume is
so small that one has to go to much larger volumes in order to do it
in a reliable way.\footnote{It has been argued that
one should not use the ``geometrical method'' advocated here as a
measure of the Hausdorff dimension, but extract it from the random walk
representation of the massive propagator, since this seems to give
more ``reasonable'' values. We disagree with this point of view. The
two methods are mathematically equivalent \cite{filk} and disagreement
reflects in our opinion the fact
that it is not possible to extract the Hausdorff
dimensions with the desired precision.}

We get a nice representation of the change in geometry between the
two phases by showing the actual distribution of geodesic length
in the universes. This is done in fig.\ 9 for the $d_4$ distances
for $h=0$. If space-time has a fractal structure with some
Hausdorff dimension $d_h$ the distribution should be like
\beq{*3.2}
P(d) \sim d^{d_h-1}
\eeq
In fig.\ 9 we have shown four curves which correspond to $\k_2 = 0.9, 1.0,
1.1$ and $1.2$ in
the critical region.
Since the transition is smooth and extrapolates from large to small
$d_h$ it is of course possible to find a $\k_2$ in the transition region where
we get a curve quite similar to \rf{*3.2} with $d_h \approx 4$ which
is of course amusing, but we do not consider the value as especially well
determined by the numerical simulations.

Finally the difference in geometry between the highly connected
phase for $\k_2$ below the critical region and very extended phase
above the critical region is reflected  in the curvature {\it distribution}.
In fig.\ 10 we have shown the distribution of the average curvature per
simplex,
defined as an average over all ten triangles forming a 4d simplex. In addition
we plot the values obtained by blocking the value of curvature over larger and
larger regions (3 and 5) in the $d_4$ distance. The value of $h$ is $0$
and the $\k_2$ values are chosen in the neighborhood of the phase
transition. The same qualitative behaviour is observed for larger values
of $h$. In all the cases the distribution seems to consist of two
parts: one sharply peaked at small curvature values and the second
rather broad, shifted towards the positive curvature.
In the highly connected phase the broad part disappears after blocking
and already after
one or two steps the distribution
approaches a $\delta$-function. {\it On larger scales we
simply have a space with constant curvature $R$}. In the other phase
it is the peaked part, which disappears after blocking and it seems
as if the distribution approaches some non-trivial limit.

\vspace{12pt}

One  disturbing aspect of our results if we want to give the above mentioned
phase transition a continuum interpretation is the fact that $\aR_c >0 $.
If we take this result literally it is difficult to attribute any
sensible naive continuum scaling to the system. If we introduce a scaling
parameter $a$, which conveniently can be identified with the link length
in the triangulation the simplest scaling would be one where
$a \to 0$ while the volume $a^4 N_4$ was kept fixed. One would then
expect the following relation between the ``bare'' lattice curvature
and the continuum curvature:
\beq{*3.4}
\aR (lattice) =  R (continuum) a^2
\eeq
which shows that if $R(continuum)$ should remain finite in the scaling limit
$a \to 0$ $\aR$ {\it must} scale to zero. It does not. We have found no way
to repair this and although it is possible to find a scaling
\beq{*3.5}
\left|\langle{R(\k_2,h)}\rangle- R_c(\k_2^c,h) \right| \sim
\left| \k_2 -\k_2^c\right|^{\delta -1}
\eeq
its significance is not clear to us due to $R_c >0$. One possible
explanation is that our formalism does not admit a tessellation of
flat space and that this in some way reflects itself in an
expectation value of $R$ ? Another possibility is, as mentioned above,
that the observed phase transition is in fact higher order and the naive
scaling relations need not hold although it is very difficult to imagine
an alternative scenario which would correspond to a physical scaling.

The appearance of $R_c > 0$ for $h=0$ was one of the motivations
to look at theories with higher curvature terms. The other motivation
was to investigate the question of universality in the spirit outlined
by Weinberg, as explained in the introduction. However, our results are
negative in the sense that even if $R_c$ indeed decreases with increasing $h$,
it does not go to zero and for not too large values of $h$ we clearly are
in the same universality class as for $h=0$. Distributions, the nature
of the transition etc.\ seem to be the same except for a displacement
in $\k_4$ and $\k_2$. As mentioned before,
for $h > 12$ we see a qualitative
change in the distributions, which might indicate a different transition.
This point requires however further investigation.

\newsection{Discussion}

The notation of a ``hot''  and a ``cold'' phase in quantum gravity
in  $d = 3$ was introduced in \cite{av2}. In the hot phase the
large entropy of ``quantum'' universes was dominant. These quantum
universes were characterized by a large  Hausdorff dimension and
a high connectivity and the hot phase
 was continuously connected to ``anti-gravity'',
where the (bare) gravitational coupling constant is negative.
In the cold phase extended structures dominated.
In fact the Hausdorff dimension seemed  close to one, suggesting
some kind of linear structure. This phase was interpreted as representing
the dominance of the conformal mode. In the regularized theory the
action is not unbounded from below,
but instead some lattice configurations
which are pure artifacts without any connection to the continuum will
dominate. These were the extended structures observed in three dimensions.
The interesting question was asked, whether it was possible at the
transition point to have truly extended structures, relevant for
continuum physics. In \cite{abkv} it was shown that the transition
in three dimensions was of first order, and a continuum limit
was ruled out from this point of view.

{}From a superficial point of view the situation does not look so different
in four dimensions. We have two phases which we again can call
hot and cold. The hot phase is continuously connected to the ``anti-gravity''
region where the bare gravitational coupling constant is negative.
As in three dimensions the cold phase is
characterized by an almost linear,
extended structure, while the hot phase has a larger Hausdorff dimension
and much larger connectivity.
In the hot phase
the average order of vertices is much
larger and the average curvature changes from being large positive in the
cold phase to small positive or even negative in the hot phase.
However, the nature  of the transition seems different in four dimensions.
We have not seen
any true hysteresis, but there are very long thermalization times
in the cold phase where the linear structures developed.
This is in contrast to three dimensions where a very pronounced
hysteresis was observed \cite{abkv}. Our data are not incompatible with
a second order transition, and this opens for the possibility that
a continuum limit can be associated with the transition. The scenario is
from this point of view quite nice: In three dimensions a first order
transition rules out a continuum limit, but we do not really want the
continuum limit in a usual sense in three dimensions since we would
be confronted with the embarrassing question of a three dimensional
graviton. The physical Hilbert space of {\it pure} three-dimensional
quantum gravity is most likely finite dimensional
\footnote{The situation might be different if we include matter fields.}
and does not allow for true dynamical fields.
The situation in four dimensions is probably very
different and it is interesting that the discretized model seems
to hint at such a difference.

It is also  encouraging  that our data share some similarities with the
results obtained by Regge calculus (\cite{hamber1} and references therein).
As explained in the introduction the philosophy of the two methods
is quite different and it would be a strong argument in favour of
universality if one manages to obtain the same results by the two
methods. In Regge calculus one also observes the two phases, and the
phase with large positive curvature is characterized by very singular
spiky configurations. They seem similar to our linear structures,
which however cannot arise by single points moving away from the
rest, as is the case in the Regge formalism. The ``hot'' phase
is in the Regge formalism characterized by a small negative
curvature which however (contrary to our results) scales
to zero at the critical point. The latest results \cite{hamber1} indicate that
one actually has a {\it first} order transition for $h=0$ and only
for a finite $h > 0$ it changes to a second order transition.
This transition for finite $h$ might have some similarity with the
change we have seen for large $h$, but we still have $R_c > 0$.
We postpone the discussion of this point to later publication.

At this point we should emphasize again that
{\it we see one major obstacle to taking a continuum limit
at the critical point and that is the fact that the average curvature
does not scale to zero}. As is seen from the fig.1\  we have
$\langle R \rangle \approx 0$ for $N_4$ large in the case where
there is no gravitational coupling constant.
This is a nice result as it implies that the measure
$\sum_{T \sim S^4}$ selects positive and negative curvature with
the same weight, as already noticed.
As soon as we add a gravitational coupling constant
$\k_2$ we get an expectation value $\langle R \rangle= R_0 (\k_2)$, which
is essentially a linear function of $\k_2$ for $\k_2$ not too large.
Since the only interesting critical behaviour takes place for $\k_2 >0$
we need at least a reinterpretation of the scaling limit in order to
be able to claim we can make contact with continuum physics. One possibility
is that the expectation value of $R_c$ has its root in the missing
tessellation of flat space, but we feel that would be a surprise in a
quantum theory of gravity. A more radical point of view would be that
there is no four-dimensional quantum theory of gravity.
Maybe quantum gravity needs matter, as the quantum theory of matter
might need gravity ?

\vspace{12pt}

{\bf Acknowledgement} It is a pleasure to thank Jeff Greensite for his
continuous, constructive criticism. One of us (J.J.) thanks the Niels Bohr
Institute for a hospitality during his stay there.


\noindent {\large \bf Figure Captions}

\vspace{12pt}

\begin{itemize}
\item[Fig.\ 1] The average curvature $\aR (\k_2)$ for $h = 0$ and $N_4 = 4000
(\bigcirc )$, $N_4 =9000 (\triangle )$, $N_4 = 16000 (+)$ and $N_4 = 32000
(\times )$.
\item[Fig.\ 2] The average curvature $\aR (\k_2)$ for $h = 0,10$ and $20$ and
$N_4 = 16000$.
\item[Fig.\ 3]
Susceptibility $\chi (\k_2)$ for $h = 0$ and $20$ and $N_4 = 4000
(\bigcirc ), N_4 = 9000 (\triangle )$ and $N_4 = 16000 (\times )$.
\item[Fig.\ 4a] $\aRR - \aR^2$ as a function of $\k_2$ for $h = 0, 10$ and $20$
and $N_4 = 4000 (\bigcirc ), 9000 (\triangle )$, $16000 (+ )$ and $32000
(\times )$.                                                         .
\item[Fig.\ 4b]  $\aRR - \aR^2$ as a function of $\k_2$ for $h = 0$ and
$N_4 = 4000 (\bigcirc ), 9000 (\triangle ), 16000 (+ )$ and $32000
(\times )$.
\item[Fig.\ 5] Critical line $\k_2^c(h)$.
\item[Fig.\ 6] Average curvature at the phase transition for $N_4 = 4000
(\bigcirc ), 9000 (\triangle )$ and $16000 (\times )$.
\item[Fig.\ 7] Average $d_1$ distance for $h = 0$ and $20$ and for $N_4 = 4000
(\bigcirc ), 9000 (\triangle )$ and $ 16000 (\times )$.
\item[Fig.\ 8] Average $d_4$ distance for $h = 0$ and $20$ and for $N_4 = 4000
(\bigcirc ), 9000 (\triangle )$ and $ 16000 (\times )$.
\item[Fig.\ 9] Distribution of the $d_4$ distances for $h = 0$ and
$\k_2 = 0.9, 1.0, 1.1$ and $1.2$.
\item[Fig.\ 10] Distribution of the curvature per simplex and blocked at a
$d_4$ distance 3 and 5.
\end{itemize}

\end{document}